\documentstyle[preprint,tighten,eqsecnum,aps]{revtex}
\begin{document}
\preprint{TRI-PP-94-64}
\draft
\title{Field transformations and the classical equation of motion in
chiral perturbation theory}
\author{S.\ Scherer\thanks{Address after Sept.\ 1, 1994:
Institut f\"ur Kernphysik, Johannes Gutenberg--Universit\"at,
J.\ J.\ Becher--Weg 45, D--55099 Mainz, Germany.}\, and H.\ W.\ Fearing}
\address{TRIUMF, 4004 Wesbrook Mall, Vancouver, B.\ C.,\\ Canada V6T 2A3}
\date{August 15, 1994}
\maketitle
\begin{abstract}
The construction  of effective Lagrangians commonly involves the application
of the `classical equation of motion' to eliminate redundant structures and
thus generate the minimal number of independent terms.
We investigate this procedure in the framework of chiral perturbation
theory.
The use of the 'classical equation of motion' is interpreted in terms of
field transformations.
Such an interpretation is crucial if one wants to bring a given Lagrangian
into a canonical form with a minimal number of terms.
We emphasize that the application of field transformations not only
eliminates structures, or, what is equivalent, expresses certain structures
in terms of already known different structures,
but also leads to a modification of coefficients of higher--order terms.
This will become relevant, once one considers effective interaction terms
beyond next--to--leading order, i.e., beyond $O(p^4)$.
\end{abstract}
\narrowtext
\section{Introduction}
Effective field theory calculations have been used in different areas of
particle physics.
Applications include chiral perturbation theory
\cite{Weinberg1,Gasser1,Gasser2}
where the approximate chiral symmetry of the underlying $QCD$ Lagrangian is
mapped into a systematic momentum expansion of the interaction between the
low--energy degrees of freedom, namely the Goldstone bosons of spontaneous
chiral symmetry breaking.
An extension of chiral perturbation theory which includes the low--lying
baryon states has been performed by several authors
\cite{Gasser3,Weinberg2,Weinberg3}.
Another example involves the treatment of the weak interaction in processes
where the external momenta are much smaller than the mass of the $W$ boson
(see Ref.\ \cite{Simma} and references therein).
Under these circumstances S--matrix elements may be expanded in terms of the
ratio $p^2/M^2_W$ and an effective Lagrangian approach is convenient to
represent such an expansion.
A further application is concerned with the effects on observables at
presently available energies of heavy particles beyond the standard model
\cite{Georgi,Arzt}.

All the above examples have in common that at a certain point in
the construction of the most general structures of the effective
Lagrangian use of the `classical equation of motion' is made in
order to eliminate redundant structures (see, e.g.,
Refs.\ \cite{Gasser1,Gasser2,Buchmueller,Pich}).
A thorough discussion on the formal aspects including, in particular,
the relation between the `classical equation of motion' and field
transformations can be found in Refs.\
\cite{Simma,Georgi,Arzt,Leutwyler,Grosse}.
The basic motivation for the use of field transformation is the
equivalence theorem which states that observables should not
depend on the specific choice for the interpolating field
\cite{Haag,Kamefuchi,Coleman}.

In this work we concentrate on {\em practical} aspects of applying field
transformations in the framework of chiral perturbation theory.
Concerning the formal justification of using field transformations in
{\em effective} theories we refer the reader to Refs. \cite{Arzt,Grosse}.
Firstly, we identify the relevant properties of field transformations
for the specific case of chiral perturbation theory.
Then we make contact with the use of the `classical equation of motion'
as a means to eliminate redundant structures in the construction of the most
general chiral Lagrangian.
Part of this discussion has already been outlined in Ref.\ \cite{Leutwyler}.

Secondly, we address the question how to bring a {\em given} effective
Lagrangian into a canonical form, i.e., a form with the minimal number
of independent terms.
This issue is of practical importance since an effective Lagrangian obtained
from, say, the bosonization of an effective quark model action, is in general
not in a canonical form.
In order to allow for an easy comparison between the predictions of different
models it is most convenient to bring the effective Lagrangian into a generally
accepted canonical form.
For example, at $O(p^4)$ the commonly used standard is the one by Gasser and
Leutwyler \cite{Gasser2}.

The formal procedure turns out to be very similar to the Foldy\---Wouthuysen
transformation \cite{Foldy,Fearing}
where a sequence of transformations is applied in order to bring
a Dirac Hamiltonian into a block--diagonal form up to a given order
in the $1/M$ expansion.
The analogy in chiral perturbation theory is a sequence of transformations
which removes redundant terms up to a given order in the momentum expansion.
In both methods it is crucial to consistently keep all terms up to
and including the desired final order at each step in the iteration process.

\section{The chiral lagrangian and field transformations}
In the momentum expansion a chiral Lagrangian is written as
a sum of terms with an increasing number of covariant derivatives, external
fields (including quark mass terms), and field strength tensors,
\begin{equation}
\label{cl}
\tilde{\cal L}={\cal L}_2+\tilde{\cal L}_4+\tilde{\cal L}_6+\dots.
\end{equation}
For the moment, we assume that the chiral Lagrangian is not yet in a {\em
canonical form}, which we define to be a form with a minimal number of
independent structures at a given order.
We denote a {\em non--canonical form} by $\tilde{\cal L}$.
The lowest--order term in Eq.\ (\ref{cl}) is, up to a total derivative,
unique \cite{Gasser2} and we may thus omit the tilde,
\begin{equation}
\label{l2}
{\cal L}_2(U) = \frac{F_0^2}{4} Tr \left ( D_{\mu} U (D^{\mu}U)^{\dagger}
\right)
+\frac{F_0^2}{4} Tr \left ( \chi U^{\dagger}+ U \chi^{\dagger} \right ),
\quad U(x)=\exp\left( i\frac{\phi(x)}{F_0} \right ),
\end{equation}
where $\phi(x)$ is a traceless, hermitian $3\times 3$ matrix containing
the eight Goldstone bosons ($\pi,K,\eta$), $D_\mu$ is a covariant derivative,
$F_0$ is the pion decay constant in the chiral limit, and $\chi$, eventually,
contains the quark masses (for details see Ref.\ \cite{Gasser2}).
The most general canonical Lagrangian at $O(p^4)$,
${\cal L}_4^{GL}$, was obtained by Gasser and Leutwyler \cite{Gasser2}.
Use of the `classical equation of motion', i.e., the one obtained from
${\cal L}_2$, was made to eliminate two
additional structures at $O(p^4)$.
In the following we will interpret this procedure in terms of field
transformations \cite{Georgi,Leutwyler,Kamefuchi,Coleman}.

Suppose we want to perform a field transformation where we
demand the same properties of another `interpolating field' $U'(x)$ as of
$U(x)$, namely,
both are $SU(3)$ matrices and transform under the group
($G=SU(3)_L\times SU(3)_R$),
parity, and charge conjugation, respectively, according to
\begin{equation}
\label{vprop}
W\stackrel{G}{\rightarrow}V_R W V_L^\dagger,\quad
W(\vec{x},t)\stackrel{P}{\rightarrow}W^\dagger(-\vec{x},t),\quad
W\stackrel{C}{\rightarrow}W^T,
\end{equation}
for $W=U$ or $U'$. In Eq.\ (\ref{vprop}) $V_L$ and $V_R$ are independent
$SU(3)$ matrices representing group elements of $G$.
Since both, $U$ and $U'$, are supposed to be $SU(3)$ matrices, they can be
related through
\begin{equation}
\label{uvrel}
U=\exp(iS(U'))U',
\end{equation}
where $S=S^{\dagger}$, and $Tr(S)=0$.
The generator $S$ may, in general, depend on all building blocks which are
required for the construction of the chiral Lagrangian.
The properties of $U$ and $U'$, Eq.\ (\ref{vprop}), result in the following
requirements for $S$:
\begin{equation}
\label{Sprop}
S\stackrel{G}{\rightarrow}V_R S V_R^\dagger,\quad
S(\vec{x},t)\stackrel{P}{\rightarrow}-U'^\dagger(-\vec{x},t)
S(-\vec{x},t)U'(-\vec{x},t),\quad
S\stackrel{C}{\rightarrow}(U'^\dagger S U')^T.
\end{equation}
For example, two such generators exist at $O(p^2)$ in the momentum expansion
(denoted by the subscript 2 of $S$)\footnote{When dealing with
$U(3)$ matrices, $S_2$ does not have to satisfy the constraint $Tr(S_2)=0$,
and thus one obtains $i\alpha_3 Tr(\chi U'^\dagger-U' \chi^\dagger)$ as
an additional independent generator.}:
\begin{equation}
\label{s2}
S_2(U')=i\alpha_1(D^2U'U'^{\dagger}-U'(D^2U')^{\dagger})
+i\alpha_2\left(\chi U'^{\dagger}-U' \chi^{\dagger}-\frac{1}{3}
Tr(\chi U'^{\dagger}-U' \chi^{\dagger})\right),
\end{equation}
where $\alpha_1$ and $\alpha_2$ are arbitrary real parameters.
Note that $S_2(U')$ depends, in addition to $U'$ and $U'^\dagger$, on
$D^2 U', \chi,\dots$.
We are somewhat sloppy since we denote this dependence collectively
by the symbol $U'$.

Let us now investigate the change in the functional form of $\tilde{\cal L}$
due to the field transformation.
After inserting Eq.\ (\ref{uvrel}) into
Eq.\ (\ref{cl}) the result can be expanded as
\begin{equation}
\label{clv}
\tilde{\cal L}(U)=\sum_{n=0}^{\infty}\delta^{(n)}\tilde{\cal L}(U',S),\quad
\delta^{(0)}\tilde{\cal L}(U',S)\equiv \tilde{\cal L}(U'),
\end{equation}
where the superscripts denote the power of S (or $D_\mu S, \dots$).
Furthermore, each term of Eq.\ (\ref{clv}) may be decomposed according to
its origin in the momentum expansion,
\begin{equation}
\label{dnl}
\delta^{(n)}\tilde{\cal L}(U',S)=\delta^{(n)}{\cal L}_2(U',S)
+\delta^{(n)}\tilde{\cal L}_4(U',S)+\dots.
\end{equation}
For example, the lowest--order change is given by\footnote{In the following
we deliberately omit total--derivative terms, since they
do not modify the equation of motion.}
\begin{equation}
\label{dell2}
\delta^{(1)}{\cal L}_2(U',S)=\frac{F^2_0}{4}
Tr\left(iS{\cal O}^{(2)}_{EOM}(U')\right),
\end{equation}
where\footnote{We adhere to the commonly used convention for the classical
equation of motion (see Eq.\ (5.9) of Ref.\ \cite{Gasser2}), even though,
from a formal point of view, it is more natural to use the hermitian
combination $i\frac{F^2_0}{4}{\cal O}^{(2)}_{EOM}$ obtained from the variation
of ${\cal L}_2$.}
\begin{equation}
\label{eom2}
{\cal O}^{(2)}_{EOM}(U')=D^2U' U'^{\dagger}-U'(D^2U')^{\dagger}
-\chi U'^{\dagger}+U'\chi^{\dagger}
+\frac{1}{3}Tr\left(\chi U'^{\dagger}-U'\chi^{\dagger}\right).
\end{equation}
In the following we do {\em not} assume that Eq.\ (\ref{eom2}) is identical to
zero, even though it has the functional form of the 'classical equation of
motion' derived from Eq.\ (\ref{l2}) \cite{Gasser2}.

For a generator $S_{2m}$ $(m\ge 1)$ of $O(p^{2m})$, the expression
$\delta^{(n)}\tilde{\cal L}_{2k}(U',S_{2m})$ is of $O(2k+2m\times n)$ in the
momentum expansion.
In practical applications one only considers the chiral Lagrangian
up to a small order, say $O(p^4)$ or $O(p^6)$.
This automatically restricts the construction of relevant generators
to $O(p^2)$ or $O(p^2)$ and $O(p^4)$, respectively.

By choosing suitable field transformations it is possible to `eliminate'
structures at $O(p^4)$ and higher in the momentum expansion, and
bring $\tilde{\cal L}$ into a canonical form $\cal L$.
To illustrate this method we first give the most general non--canonical form of
the chiral Lagrangian at $O(p^4)$ (prior to the application of the
`classical equation of motion') \cite{Rudy},
\begin{eqnarray}
\label{l4mg}
\tilde{\cal L}_4(U)&=&{\cal L}_4^{GL}(U)
+c_1 Tr\left((D^2U U^\dagger-U (D^2U)^\dagger){\cal O}^{(2)}_{EOM}\right)
\nonumber\\&&
+c_2 Tr\left((\chi U^\dagger-U\chi^\dagger){\cal O}^{(2)}_{EOM}\right),
\end{eqnarray}
where we do not need to present the explicit form of ${\cal L}_4^{GL}$ of
Gasser and Leutwyler for the present purpose.
If we insert the field transformation of Eq.\ (\ref{uvrel}) with $S_2$
of Eq.\ (\ref{s2}) into ${\cal L}_2(U)+\tilde{\cal L}_4(U)$, and choose
$\alpha_1=4c_1/F^2_0$ and $\alpha_2=4 c_2/F^2_0$, the last two terms
of Eq.\ (\ref{l4mg}) and the modification $\delta^{(1)}{\cal L}_2$
precisely cancel and one obtains the result of Gasser and Leutwyler
(after renaming $U'$ into $U$).
In other words, the Lagrangians\footnote{For convenience we use the same symbol
$U$ in both Lagrangians.} ${\cal L}_2(U)+{\cal L}_4^{GL}(U)$
and ${\cal L}_2(U)+\tilde{\cal L}_4(U)$ are equivalent up to $O(p^4)$ and thus,
according to the equivalence theorem, should generate the same (on--shell)
S--matrix elements at $O(p^4)$.
In this sense, namely by identifying structures which are proportional
to ${\cal O}^{(2)}_{EOM}$, the `classical equation of motion' can be
used to eliminate terms in an effective Lagrangian.

This situation is often interpreted in the following manner
(see, e.g., Refs.\ \cite{Buchmueller,Donoghue}).
The equation of motion in the presence of higher--order terms can
be written as
\begin{equation}
\label{eomhor}
{\cal O}^{(2)}_{EOM}(U)=0+O(p^4).
\end{equation}
Inserting Eq.\ (\ref{eomhor}) into Eq.\ (\ref{l4mg}) and dropping
terms which are of $O(p^6)$ generates ${\cal L}_4^{GL}(U)$.
However, even though this argument is intuitively appealing and correct
at this order, it also contains a potential source of error, as soon as one
tries to apply it in the presence of higher--order structures.
We will discuss this point in more detail below.

It is straightforward to extend the use of field transformations to
higher--order terms.
Suppose the higher--order Lagrangian contains a chirally invariant
structure of the type
\begin{eqnarray}
\label{ftho}
\lefteqn{Tr\left( (D^2 U U^\dagger-U (D^2 U)^\dagger)A\right)=
 Tr\left((D^2 U U^\dagger-U (D^2 U)^\dagger)\bar{A}\right)}\nonumber\\
&&= Tr\left({\cal O}^{(2)}_{EOM}\bar{A}\right)
+Tr\left(A(\chi U^\dagger-U \chi^\dagger)\right)
 -\frac{1}{3}Tr(\chi U^\dagger-U\chi^\dagger)Tr(A),
\end{eqnarray}
where $\bar{A}=A-\frac{1}{3}Tr(A)$ is traceless and of $O(p^4)$ or higher.
We used $Tr(D^2 U U^\dagger - U (D^2 U)^\dagger)=0$ and $Tr(\bar{A})=0$ in the
first and second step, respectively.
With the help of a field transformation involving $\bar{A}$ the first
term can be eliminated and the application of the field transformation
looks as if one had simply made the replacement
\begin{equation}
\label{rep}
D^2U U^\dagger-U (D^2 U)^\dagger \rightarrow \chi U^\dagger- U \chi^\dagger
-\frac{1}{3} Tr\left(\chi U^\dagger - U \chi^\dagger\right),
\end{equation}
in order to replace the original expression in terms of two structures
which one anyway has to consider\footnote{At present
we cannot exclude the possibility that in some cases it might be
advantageous to reverse the direction of the arrow in Eq.\ (\ref{rep}),
i.e., to perform the replacement the other way around.}.

However, at this point one should carefully distinguish between the following
two situations which are of practical interest.
The first regards the pure {\em identification} of terms necessary
for the construction of the most general chiral Lagrangian.
For this purpose it is completely sufficient to use the `classical
equation of motion' in the sense of Eq.\ (\ref{rep}).
This can be understood as follows \cite{Georgi}.
Even though a suitable field transformation containing a generator of
$O(p^{2m})$ not only removes or reexpresses a term at $O(p^{2m+2})$ but
also modifies the interaction terms at $O(p^{2m+4})$ and higher, this
modification concerns the coefficients of structures which one anyway has
to consider in the construction of the most general Lagrangian.
As long as one is not yet interested in the specific values of these
coefficients, one can freely make use of Eq.\ (\ref{rep}) to reduce
the number of structures.

A different and more complicated situation occurs if one actually wants to
bring a given non--canonical form of a chiral Lagrangian into a canonical form.
Non--canonical forms typically arise in effective Lagrangians
which are obtained
after bosonization of an effective $QCD$ inspired quark interaction
\cite{Balog,Ebert,Ball,Belkov}.
For the purpose of comparing different effective Lagrangians it is most
convenient to bring them into a commonly accepted standard form and to simply
compare the predictions for the coefficients.
For example, at $O(p^4)$ it is standard practice to use the canonical
form of Gasser and Leutwyler \cite{Gasser2}.

In the following we will discuss the issue of transforming chiral Lagrangians
beyond $O(p^4)$ into a canonical form.
For practical purposes we restrict our discussion to terms up to $O(p^6)$.
Clearly, an extension to higher orders is straightforward, but very tedious.
Suppose one starts with predictions for the coefficients of a Lagrangian
$\tilde{\cal L}={\cal L}_2+\tilde{\cal L}_4
+\tilde{\cal L}_6$, where $\tilde{\cal L}_4$ is not yet in the form of
the Lagrangian of Gasser and Leutwyler.
In addition, $\tilde{\cal L}_6$ may not yet be in the canonical form.

A naive application of the equation of motion argument, namely, a replacement
according to Eq.\ (\ref{rep}) will not yield the correct result,
i.e., it will not generate an equivalent Lagrangian which reproduces the
same S--matrix elements as the original Lagrangian.
The reason is that even though the replacement using Eq.\ (\ref{rep})
produces the required structures, it does not automatically give the correct
coefficients.
To be specific, the transformation procedure using $S_2$ as described
following Eq.\ (\ref{l4mg}) results in the following Lagrangian at $O(p^6)$,
\begin{equation}
\tilde{\cal L}'_6(U')=\tilde{\cal L}_6(U')+
\delta^{(1)}\tilde{\cal L}_4(U',S_2)+
\delta^{(2)}{\cal L}_2(U',S_2),
\end{equation}
where
\begin{eqnarray}
\label{dell4}
\delta^{(1)}\tilde{\cal L}_4 (U',S)&=&\frac{F^2_0}{4}
Tr\left(iS{\cal O}^{(4)}_{EOM}(U')\right),\nonumber\\
\delta^{(2)}{\cal L}_2(U',S)&=&\frac{F^2_0}{4} Tr\left(S(D_\mu U' U'^\dagger
D^\mu S- D^\mu S D_\mu U' U'^\dagger - D^2 S)\right.\nonumber\\
&&\left.-\frac{1}{2}(\chi U'^\dagger + U' \chi^\dagger) S^2\right).
\end{eqnarray}
In Eq.\ (\ref{dell4}) ${\cal O}^{(4)}_{EOM}$ refers to the contribution
to the equation of motion derived from $\tilde{\cal L}_4$.
Once again, in order to be consistent with Eq.\ (\ref{eom2}) we choose
a form for ${\cal O}^{(4)}_{EOM}$ which differs by a factor $-4i/F^2_0$ from
what one would obtain with the variational principle.
The covariant derivative of an operator $A$ which transforms under the group as
$V_R A V_R^\dagger$ (such as $S$) is defined as
\begin{equation}
\label{covdera}
D_\mu A = \partial_\mu A-iR_\mu A +iA R_\mu.
\end{equation}
Clearly, the two terms of Eq.\ (\ref{dell4}) result in an additional
contribution to the Lagrangian at $O(p^6)$ which one would have missed
had one simply applied Eq.\ (\ref{rep}) to $\tilde{\cal L}_4$,
i.e., had one simply dropped the terms proportional to the classical
equation of motion.
Such a procedure is only correct up to $O(p^4)$.
Furthermore, from Eq.\ (\ref{dell4}) one can also deduce that even a
generalization of Eq.\ (\ref{eomhor}) taking into account
${\cal O}^{(4)}_{EOM}$ resulting from $\tilde{\cal L}_4$,
\begin{equation}
\label{eomhor2}
{\cal O}^{(2)}_{EOM}=-{\cal O}^{(4)}_{EOM}+O(p^6),
\end{equation}
does not yield an equivalent Lagrangian since one misses the contribution of
$\delta^{(2)}{\cal L}_2$.
Even though Eq.\ (\ref{eomhor2}) is a straightforward extension of
Eq.\ (\ref{eomhor}), it obviously cannot be used to rewrite structures
in $\tilde{\cal L}_4$ with a precision up to and including $O(p^6)$.
At first sight this looks surprising as one {\em appears} to use the
same method as before.
However, there is a fundamental difference between applying Eq.\ (\ref{eomhor})
to Eq.\ (\ref{l4mg}) and the corresponding replacement using
Eq.\ (\ref{eomhor2}).
In the first case information of ${\cal L}_2$ is used in
the {\em higher--order}
term $\tilde{\cal L}_4$, whereas in the second case one tries to put back
`information' derived from $\tilde{\cal L}_4$ into the {\em same} term.
We will illustrate this point in a simple toy model \cite{Simma} in the
Appendix.

We now describe a systematic procedure which consists of successively applying
transformations with generators where the order in the momentum expansion
increases with each iteration.
Let $2N$ denote the maximal order of a given non--canonical chiral Lagrangian
\begin{equation}
\label{l2n}
\tilde{\cal L}= {\cal L}_2+\tilde{\cal L}_4 + \dots +\tilde{\cal L}_{2N}
\end{equation}
which we want to bring into a canonical form up to the same order.
Prior to the first iteration step one brings, by suitably adding
and subtracting
terms, the $O(p^4)$ Lagrangian $\tilde{\cal L}_4$ into the form of
Eq.\ (\ref{l4mg}) and identifies the coefficients $c_1$ and $c_2$.
The first transformation
\begin{equation}
\label{ft}
U=\exp(iS_2(U^{(1)}))U^{(1)}
\end{equation}
with $S_2$ given in Eq.\ (\ref{s2}) and $\alpha_1=4c_1/F_0^2$ and
$\alpha_2=4c_2/F_0^2$,
results in an equivalent effective Lagrangian which is canonical up to
$O(p^4)$.
Before applying the second iteration one has to manipulate the new
$\tilde{\cal L}'_6$ obtained after the first transformation in such a manner
as to obtain as many terms as possible containing ${\cal O}^{(2)}_{EOM}$
(see Eq.\ (\ref{ftho})).
One then applies the transformation
\begin{equation}
\label{st}
U^{(1)}=\exp(iS_4(U^{(2)}))U^{(2)},
\end{equation}
with a suitable choice of $S_4$.
The resulting Lagrangian is canonical up to $O(p^6)$.
Clearly, one has to perform $N-1$ iterations
in order to obtain a canonical Lagrangian up to $O(p^{2N})$.
At each step it is essential to consistently keep all the terms
$\delta^{(n)}\tilde{\cal L}_{2k}(U^{(m)},S_{2m}(U^{(m)}))$
with $N =k+m\times n$
where $U^{(m)}$ denotes the interpolating field after
the $m$th transformation.
Only at the final iteration step is it allowed to simply perform the
replacement according to Eq.\ (\ref{rep}).
The reason is that the last transformation only removes terms which are
proportional to the 'classical equation of motion' and modifies higher--order
terms which one is not interested in.
This is essentially the reason why the simple application of Eq.\ (\ref{rep})
to the $O(p^4)$ Lagrangian led to the correct answer up to $O(p^4)$.

We note the similarity of the algorithm to the
Foldy--\-Wouthuysen method \cite{Foldy,Fearing} which is used to transform
a Dirac Hamiltonian to a block--diagonal form up to a finite order
$n$ in $1/M$.
It is well--known from this method that at each iteration step one
consistently has to keep all terms up to order $n$ in the interaction
Hamiltonian.
This is required to generate S--matrix elements which are equivalent
up to order $n$ in the $1/M$ expansion.

\section{Summary}
We considered the use of field transformations for eliminating
redundant structures in the framework of chiral perturbation theory.
After stating the formal requirements for field transformations we
derived the change in the functional form of the effective Lagrangian.
In particular, we showed that for the purpose of identifying redundant
terms, it is sufficient to make use of a simple replacement
involving the `classical equation of motion'.
However, we explicitly demonstrated that this procedure is, in general,
not suited to transform a given non--canonical Lagrangian into
a canonical form.
The reason is that it does not provide the quantitative change of
the coefficients which multiply the independent structures.
We described an iterative algorithm similar to the Foldy--Wouthuysen method
which transforms a non--canonical into a canonical form up to any
desired order in the momentum expansion.

\section{Acknowledgements}

This work was supported in part by a grant from the Natural Sciences and
Engineering Research Council of Canada. One of the authors (S.\ S.) would
like to thank A.\ Schaale, A.\ A.\ Bel'kov and A.\ V.\ Lanyov for
stimulating discussions.

\begin{appendix}

\section{Toy model}
As an instructive example let us consider the following toy model \cite{Simma}.
The lowest--order Lagrangian ${\cal L}_0$ is taken to be the sum of two free
Lagrangians of a massive scalar particle and a massless fermion, respectively,
\begin{equation}
\label{l0tm}
{\cal L_0}=\frac{1}{2}\left(\partial_\mu\Phi\partial^\mu\Phi-m^2\Phi^2\right)
+i\bar{\Psi}\partial\hspace{-.5em}/\Psi.
\end{equation}
We introduce an {\em effective} interaction describing the decay of the scalar
particle into a fermion--antifermion pair,
\begin{equation}
\label{leffin}
{\cal L}_{eff}=g_{eff} \bar{\Psi}\Psi\Box\Phi,
\end{equation}
where the coupling constant $g_{eff}$ has the dimension $energy^{-2}$.
For example, with the above Lagrangian the invariant amplitude for
fermion--fermion scattering,
$A(p_a)+B(p_b) \rightarrow A(p_a')+B(p_b')$, is found to be
\cite{Simma}
\begin{equation}
\label{ia}
M=-ig^2_{eff}\left(\frac{t^2}{t-m^2}+\frac{u^2}{u-m^2}\right),
\end{equation}
where $t=(p_a-p_a')^2=(p_b-p_b')^2$ and $u=(p_a-p_b')^2=(p_b-p_a')^2$
are the Mandelstam variables \cite{Itzykson}.
Of course, the effective Lagrangian will contain interaction terms being of
higher than first order in $g_{eff}$, which may contribute to the scattering
amplitude.
However, for the point we want to make, a knowledge of the explicit form of
such terms is not required.

Now let us consider the transformation method. If we introduce
$\Phi=\chi+\delta\chi$, the Lagrangian may be written as
\begin{eqnarray}
\label{tl}
{\cal L} (\Phi,\Psi)={\cal L}' (\chi,\Psi)&=&
\frac{1}{2}\left(\partial_\mu\chi\partial^\mu\chi-m^2\chi^2\right)
+i\bar{\Psi}\partial\hspace{-.5em}/\Psi
-\delta\chi(\Box+m^2)\chi\nonumber\\
&&-\frac{1}{2}\delta\chi\Box\delta\chi-\frac{1}{2}m^2
(\delta\chi)^2+g_{eff}\bar{\Psi}\Psi\Box(\chi+\delta\chi),
\end{eqnarray}
where we dropped an irrelevant total derivative. For the choice
$\delta\chi=g_{eff}\bar{\Psi}\Psi$ the terms $-\delta\chi\Box\chi$ and
$g_{eff}\bar{\Psi}\Psi\Box\chi$ precisely cancel, and we are left with
the Lagrangian
\begin{eqnarray}
\label{ml}
{\cal L}'(\chi,\Psi)&=&{\cal L}_0(\chi,\Psi)+\tilde{\cal L}_{eff}(\chi,\Psi)
\nonumber\\
&=& \frac{1}{2}\left(\partial_\mu\chi\partial^\mu\chi-m^2\chi^2\right)
+i\bar{\Psi}\partial\hspace{-.5em}/\Psi\nonumber\\
&&-m^2g_{eff}\bar{\Psi}\Psi\chi +\frac{1}{2}g_{eff}^2
\bar{\Psi}\Psi(\Box-m^2)(\bar{\Psi}\Psi).
\end{eqnarray}
The Lagrangian of Eq.\ (\ref{ml}) is identical with the one obtained
in Ref.\ \cite{Simma} where, however, a different procedure has been
used to derive a `reduced' effective Lagrangian yielding the same
on--shell matrix elements.

Clearly, for an effective interaction term, ${\cal L}^{(2)}_{eff}(\Phi)$,
which is already of order $g^2_{eff}$, one finds
\begin{equation}
\label{}
{\cal L}_{eff}^{(2)}(\Phi)={\cal L}_{eff}^{(2)}(\chi)+O(g^3_{eff}),
\end{equation}
and thus the same contribution results from such a term at order $g^2_{eff}$
in both representations.
This is the reason why we did not explicitly specify the form of the
higher--order Lagrangian.

If one wants to investigate processes at order $g_{eff}^2$, it is essential
to consistently keep terms in the transformation up to that order.
In order to see this we consider the scattering amplitude of two fermions
at order $g^2_{eff}$ in the framework of Eq.\ (\ref{ml}).
The pole and contact diagrams yield, respectively,
\begin{eqnarray}
M_{P}&=&-ig^2_{eff}\left(\frac{m^4}{t-m^2}+\frac{m^4}{u-m^2}\right),\nonumber\\
M_{C}&=&-ig^2_{eff}\left(t+u+2m^2\right),
\end{eqnarray}
and the sum of the two contributions is the same as Eq.\ (\ref{ia}).
It is important to realize that the contact term results from two
different sources.
The first is a quadratic change in ${\cal L}_0$ and the second is a first order
change in ${\cal L}_{eff}$.
Only if one consistently keeps {\em both} terms, does one reproduce the correct
answer.

A naive replacement in Eq.\ (\ref{leffin}) using the `classical equation of
motion' ($EOM$) for $\Phi$ derived from ${\cal L}_0$,
\begin{equation}
\label{nr}
g_{eff} \bar{\Psi}\Psi\Box\Phi=g_{eff} \bar{\Psi}\Psi(\Box+m^2)\Phi
-g_{eff}m^2 \bar{\Psi}\Psi\Phi
\stackrel{EOM}{\rightarrow}-g_{eff}m^2 \bar{\Psi}\Psi\Phi,
\end{equation}
leads to incorrect results for S-matrix elements at order $g_{eff}^2$ since
it does not generate the contact terms at order $g_{eff}^2$.
One might argue that, in fact, one should use the equation of motion
derived from ${\cal L}_0+{\cal L}_{eff}$,
\begin{equation}
\label{eomeff}
\Box\Phi=-m^2\Phi+g_{eff}\Box(\bar{\Psi}\Psi),
\end{equation}
to reexpress $\Box\Phi$ in ${\cal L}_{eff}$, and then work with the
new Lagrangian. However, even though this method
generates a contact term at order $g_{eff}^2$, it does not
reproduce the correct
S-matrix for fermion--fermion scattering as given by Eq.\ (\ref{ia}).

\end{appendix}

\end{document}